\documentclass[twoside]{article}
\usepackage{qic}
\usepackage{graphicx}
\usepackage{dcolumn}
\usepackage{bm}
\usepackage{braket}
\usepackage{enumerate}
\usepackage{booktabs}

\usepackage{amssymb,amsfonts}
\usepackage{amsmath}
\usepackage{epstopdf}
\usepackage{wrapfig}
\usepackage[cp1252]{inputenc}

\usepackage[english]{babel}
\usepackage{color}

\makeatletter
\def\hlinewd#1{%
	\noalign{\ifnum0=`}\fi\hrule \@height #1 %
	\futurelet\reserved@a\@xhline}
\makeatother

\begin{document}
\setlength{\textheight}{8.0truein}    

\runninghead{FACTORING USING $2n+2$ QUBITS WITH TOFFOLI BASED MODULAR MULTIPLICATION}
            {Thomas H\"aner, Martin Roetteler, Krysta M. Svore}

\normalsize\textlineskip
\thispagestyle{empty}
\setcounter{page}{1}


\vspace*{0.88truein}

\alphfootnote

\fpage{1}

\begin{center}
	\bf FACTORING USING $\mathbf{2n+2}$ QUBITS WITH TOFFOLI BASED MODULAR MULTIPLICATION
\end{center}
\vspace*{0.37truein}
\centerline{\footnotesize
THOMAS H\"{A}NER\footnote{haenert@phys.ethz.ch}}
\vspace*{0.015truein}
\centerline{\footnotesize\it Station Q Quantum Architectures and Computation Group, Microsoft Research} 
\baselineskip=10pt
\centerline{\footnotesize\it Redmond, WA 98052, USA}
\vspace*{0.015truein}
\centerline{\footnotesize and}
\vspace*{0.015truein}
\centerline{\footnotesize\it Institute for Theoretical Physics, ETH Zurich}
\baselineskip=10pt
\centerline{\footnotesize\it 8093 Zurich, Switzerland}
\vspace*{10pt}
\centerline{\footnotesize 
MARTIN ROETTELER\footnote{martinro@microsoft.com}}
\vspace*{0.015truein}
\centerline{\footnotesize\it Station Q Quantum Architectures and Computation Group, Microsoft Research} 
\baselineskip=10pt
\centerline{\footnotesize\it Redmond, WA 98052, USA}
\vspace*{10pt}
\centerline{\footnotesize 
KRYSTA M.~SVORE\footnote{ksvore@microsoft.com}}
\vspace*{0.015truein}
\centerline{\footnotesize\it Station Q Quantum Architectures and Computation Group, Microsoft Research} 
\baselineskip=10pt
\centerline{\footnotesize\it Redmond, WA 98052, USA}
\vspace*{0.225truein}
\publisher{(received date)}{(revised date)}

\vspace*{0.21truein}

\abstracts{
We describe an implementation of Shor's quantum algorithm to factor $n$-bit integers using only $2n{+}2$ qubits. In contrast to previous space-optimized implementations, ours features a purely Toffoli based modular multiplication circuit. The circuit depth and the overall gate count are in $\mathcal O(n^3)$ and $\mathcal O(n^3\log n)$, respectively. We thus achieve the same space and time costs as Takahashi et al.~\cite{takahashi2006quantum}, while using a purely classical modular multiplication circuit. As a consequence, our approach evades most of the cost overheads originating from rotation synthesis and enables testing and localization of some faults in both, the logical level circuit and an actual quantum hardware implementation. Our new (in-place) constant-adder, which is used to construct the modular multiplication circuit, uses only dirty ancilla qubits and features a circuit size and depth in $\mathcal O(n\log n)$ and $\mathcal O(n)$, respectively.
}{}{}

\vspace*{10pt}

\vspace*{3pt}
\communicate{to be filled by the Editorial}

\vspace*{1pt}\textlineskip    
\section{\label{sec:intro}Introduction}

\begin{figure*}[!t]
	\includegraphics[width=\textwidth]{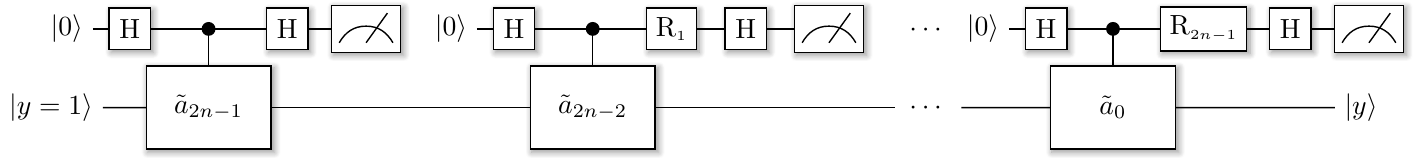}
	\fcaption{Circuit for Shor's algorithm as in \cite{beauregard2002circuit}, using the single-qubit semi-classical quantum Fourier transform from \cite{griffiths1996semiclassical}. In total, $2n$ modular multiplications by $\tilde a_i=a^{2^i}\operatorname{mod}N$ are required (denoted by $\tilde a_i$-gates in the circuit). The phase-shift gates $R_k$ are given by $\begin{tiny}\left(
		\begin{matrix}
		1&0\\
		0&e^{i\theta_k}
		\end{matrix}\right)
		\end{tiny}$ with $\theta_k=-\pi\sum_{j=0}^{k-1}2^{k-j}m_i$, where the sum runs over all previous measurements $j$ and $m_j\in\{0,1\}$ denotes the respective measurement result ($m_0$ denotes the least significant bit of the final answer and is obtained in the first measurement).}
	\label{fig:shor}
\end{figure*}

Quantum computers offer an exponential speedup over their classical counterparts for solving certain problems, the most famous of which is Shor's algorithm \cite{shor1994algorithms} for factoring a large number $N$ --- an algorithm that enables the breaking of many popular encryption schemes including RSA. At the core of Shor's algorithm lies a modular exponentiation of a constant $a$ by a superposition of values $x$ stored in a register of $2n$ quantum bits (qubits), where $n=\lceil\log_2N\rceil$. Denoting the $x$-register by $\Ket x$ and adding a result register initialized to $\Ket 0$, this can be written as
\[
	\Ket x\Ket 0\mapsto\Ket x\Ket{a^x\operatorname{mod}N}.
\]
This mapping can be implemented using $2n$ conditional modular multiplications, each of which can be replaced by $n$ (doubly-controlled) modular additions using repeated-addition-and-shift \cite{beauregard2002circuit}. For an illustration of the circuit, see Fig.~\ref{fig:shor}.

There are many possible implementations of Shor's algorithm, all of which offer deeper insight into space/time trade-offs by, e.g., using different ways of implementing the circuit for adding a known classical constant $c$ to a quantum register $\Ket a$ (see Table~\ref{tbl:adders}). The implementation given in Ref.~\cite{takahashi2006quantum} features the lowest known number of qubits and uses Draper's addition in Fourier space \cite{draper2000addition}, allowing factoring to be achieved using only $2n+2$ qubits at the cost of a circuit size in $\Theta(n^3\log n)$ or even $\Theta(n^4)$ when using exact quantum Fourier transforms (QFT). Furthermore, the QFT circuit features many (controlled) rotations, which in turn imply a large T-gate count when quantum error-correction (QEC) is required. Implementations using classically-inspired adders as in Ref.~\cite{cuccaro2004new}, on the other hand, yield circuits with as few as $3n+\mathcal O(1)$ qubits and $\mathcal O(n^3)$ size. Such classical reversible circuits have several advantages over Fourier-based arithmetic. In particular, 
\begin{enumerate}
	\item they can be efficiently simulated on a classical computer, i.e., the logical circuits can be tested on a classical computer,
   \item they can be efficiently debugged when the logical level circuit is implemented in actual quantum hardware implementation, and
	\item they do not suffer from the overhead of single-qubit rotation synthesis~\cite{KMM:2016,Selinger:2015,RS:2016,BRS:2015,BRS:2015b} when employing QEC.
\end{enumerate}

\renewcommand{\arraystretch}{1.4}
\begin{table}[t]
	\centering
	\begin{tabular}{l@{\;\;\;}c@{\;\;}c@{\;\;}c@{\;\;}c}
		& Cuccaro \cite{cuccaro2004new} & Takahashi 
\cite{takahashi2009quantum} & Draper \cite{draper2000addition} & Our adder\\\midrule\addlinespace[\belowrulesep]
\hline
		Size  & $\Theta(n)$ & $\Theta(n)$ & $\Theta(n^2)$ & $\Theta(n\log n)$\\
		Depth & $\Theta(n)$ & $\Theta(n)$ & $\Theta(n)$ & $\Theta(n)$ \\
		Ancillas & $n{+}1$ (clean) & $n$ (clean) & $0$ & $1$ dirty\\
\bottomrule\addlinespace[\belowrulesep]
	\end{tabular}
	\tcaption{Costs associated with various implementations of addition $\Ket a\mapsto\Ket{a+c}$ of a value $a$ by a classical constant $c$. Our adder uses 1 dirty ancilla to achieve a depth in $\Theta(n)$. More dirty qubits (up to $n$) allow to reduce the depth by constant factors.}
	\label{tbl:adders}
\end{table}

We construct our $\mathcal O(n^3\log n)$-sized implementation of Shor's algorithm from a Toffoli based in-place constant-adder, which adds a classically known $n$-bit constant $c$ to the $n$-qubit quantum register $\Ket a$, i.e., which implements $\Ket a\mapsto \Ket{a+c}$ where $a$ is an arbitrary $n$-bit input and $a+c$ is an $n$-bit output (the final carry is ignored).

Our main technical innovation is to obtain space savings by making use of {\em dirty} ancilla qubits which the circuit is allowed to borrow during its execution. By a dirty ancilla we mean---in contrast to a clean ancilla which is a qubit that is initialized in a known quantum state---a qubit which can be in an arbitrary state and, in particular, may be entangled with other qubits. In our circuits, whenever such dirty ancilla qubits are borrowed and used as scratch space they are then returned in exactly the same state as they were in when they were borrowed. 

Our addition circuit requires $\mathcal O(n\log n)$ Toffoli gates and has an overall depth of $O(n)$. 
Following Beauregard \cite{beauregard2002circuit}, we construct a modular multiplication circuit using this adder and report the gate counts of Shor's algorithm in order to compare our implementation to the one of Takahashi et al.~\cite{takahashi2006quantum}, who used Fourier addition as a basic building block. 

Our paper is organized as follows: in Section~\ref{sec:adder}, we describe our Toffoli based in-place addition circuit, including parallelization. We then provide implementation details of the modular addition and the controlled modular multiplier in Section~\ref{sec:mult}, where we use the same constructions as in~\cite{takahashi2006quantum,beauregard2002circuit}. We verify the correctness of our circuit using simulations and present numerical evidence for the correctness of our cost estimates in Section~\ref{sec:exp}. Finally, in Section~\ref{sec:advantages}, we provide arguments in favor of having Toffoli based networks in quantum computing.

\section{\label{sec:adder}Toffoli based in-place addition}

One possible way to construct an (inefficient) adder is to note that one can calculate the final bit $r_{n-1}$ of the result $r=a+c$ using $n-1$ borrowed dirty qubits $g$. By \textit{borrowed dirty} qubits we mean that the qubits are in an unknown initial state and must be returned to this state. Takahashi et al. hardwired a classical ripple-carry adder to arrive at a similar circuit, which they used to optimize the modular addition in Shor's algorithm \cite{takahashi2006quantum}. We construct our CARRY circuit from scratch, which allows to save $\mathcal O(n)$ NOT gates as follows.

Since there is no way of determining the state of the $g$-register without measuring, one can only use toggling of qubits to propagate information, as done by Barenco et al.~for the multiply-controlled-NOT using just one borrowed dirty ancilla qubit~\cite{barenco1995elementary}. We choose to encode the carry using such qubits, i.e., the toggling of qubit $g_i$, which we denote as $g_i=1$, indicates the presence of a carry from bit $i$ to bit $i+1$ when adding the constant $c$ to the bits of $a$. Thus, $g_i$ must toggle if (at least) one of the following statements is true:
\begin{equation*}
a_i=c_i=1, \quad g_{i-1}=a_i=1, \quad \text{or} \;\; g_{i-1}=c_i=1.
\end{equation*}
If $c_i=1$, one must toggle $g_{i+1}$ if $a_i=1$, which can be achieved by placing a CNOT gate with target $g_{i+1}$ and control $a_i=1$. Furthermore, there may be a carry when $a_i=0$ but $g_{i-1}=1$. This is easily solved by inverting $a_i$ and placing a Toffoli gate with target $g_i$, conditioned on $a_i$ and $g_{i-1}$. If, on the other hand, $c_i=0$, the only way of generating a carry is for $a_i=g_{i-1}=1$, which can be solved with the Toffoli gate from before. 

\begin{figure}[t]
	\centering
	\includegraphics[width=.7\linewidth]{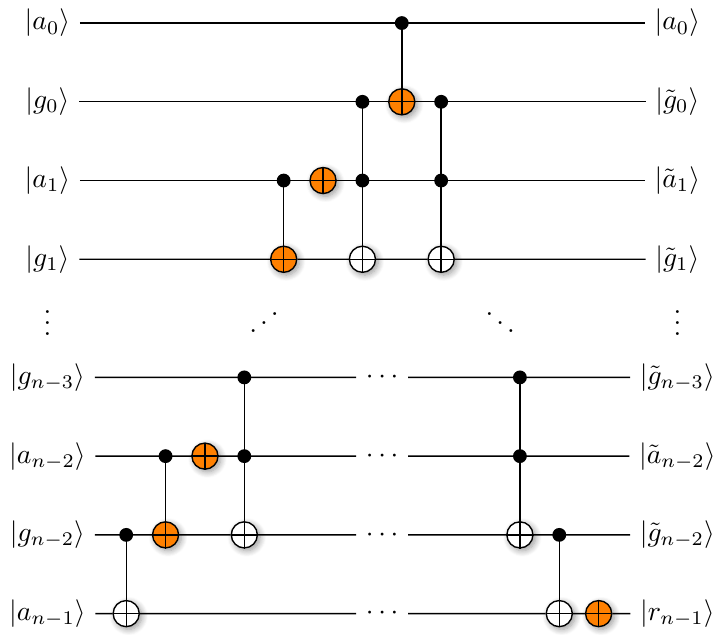}
	\bigskip
	\fcaption{Circuit computing the last bit of $r=a+c$ using dirty qubits $g$. An orange (dark) gate acting on a qubit with index $i$ must be dropped if the $i$-th bit of the constant $c$ is $0$. The entire sequence must be run again in reverse (without the gates acting on $a_{n-1}$) in order to reset all qubits to their initial value except for $r_{n-1}$.}
	\label{fig:adder1}
\end{figure}

Thus, in summary, one always places the Toffoli gate conditioned on $g_{i-1}$ and $a_i$, with target $g_i$ and, if $c_i=1$, one first adds a CNOT and a NOT gate. This classical conditioning during circuit-generation time is indicated by colored gates in Fig.~\ref{fig:adder1}. In order to apply the Toffoli gate conditioned on the toggling of $g_{i-1}$, one places it before the potential toggling, and then again afterwards such that if both are executed, the two gates cancel. Finally, the borrowed dirty qubits and the qubits of $a$ need to be restored to their initial state (except for the highest bit of $a$, which now holds the result). This is done by running the entire circuit backwards, ignoring all gates acting on $a_{n-1}$.

\begin{figure}[t]
	\centering
	\includegraphics[width=.9\linewidth]{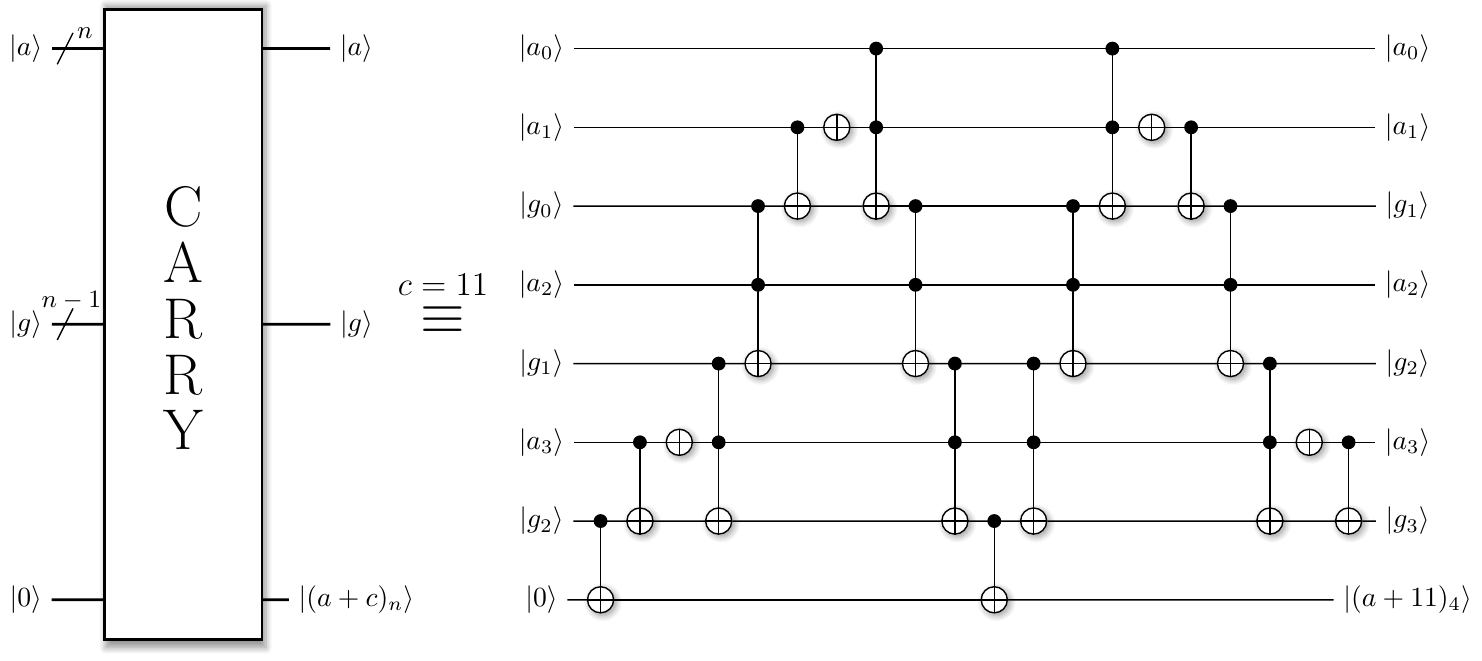}
	\bigskip
	\fcaption{Example circuit computing the final carry of $r=a+11$ derived from the construction depicted in Fig.~\ref{fig:adder1}. The binary representation of the constant $c$ is $c=11=1011_2$, i.e., the orange gates in Fig.~\ref{fig:adder1} acting on qubit index $2$ have been removed since $c_2=0$. Furthermore, the optimization mentioned in the text has been applied, allowing to remove $g_0$ in Fig.~\ref{fig:adder1}.}
	\label{fig:adderex}
\end{figure}

One can easily save the qubit $g_0$ in Fig.~\ref{fig:adder1} by conditioning the Toffoli gate acting on $g_1$ directly on the value of $a_0$ (instead of testing for toggling of $g_0$). If $c_0=0$, the two Toffoli gates can be removed altogether since the CNOT acting on $g_0$ would not be present and the two Toffolis would cancel. If, on the other hand, $c_0=1$, the two Toffoli gates can be replaced by just one, conditioned on $a_0$. See Fig.~\ref{fig:adderex} for the complete circuit computing the last bit of $a$ when adding the constant $c=11$.

If one were to iteratively calculate the bits $n-2,...,1,0$, one would arrive at an $O(n^2)$-sized addition circuit using $n-1$ borrowed dirty ancilla qubits. This is the same size as the Fourier addition circuit \cite{draper2000addition}, unless one uses an approximate version of the quantum Fourier transform bringing the size down to $\mathcal O(n\log{\frac n\varepsilon})$ \cite{cleve2000fast}. We improve our construction to arrive at a size in $\mathcal O(n\log n)$ in the next subsection.

\subsection{Serial implementation}
\begin{figure}[t]
	\centering
	\includegraphics[width=.85\linewidth]{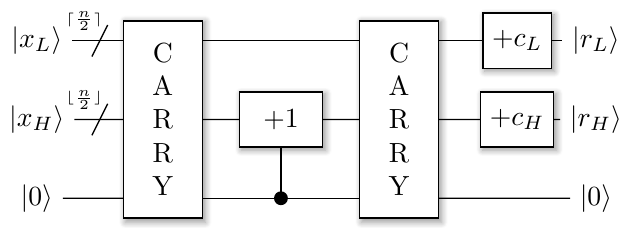}
	\bigskip
	\fcaption{Circuit for adding the constant $a$ to the register $x$. $x_H$ and $x_L$ denote the high- and low-bit part of $x$. The CARRY gate computes the carry of the computation $x_L+a_L$ into the qubit with initial state $\Ket 0$, borrowing the $x_H$ qubits as dirty ancillae. This carry is then taken care of by an incrementer gate acting on the high-bits of $x$. Applying this construction recursively yields an $\mathcal O(n\log n)$ addition circuit with just one ancilla qubit (the $\Ket 0$ qubit in this figure).}
	\label{fig:addernlogn}
\end{figure}

An $\mathcal O(n\log n)$-sized addition circuit can be achieved by applying a divide-and-conquer scheme to the straight-forward addition idea mentioned above (see Fig.~\ref{fig:addernlogn}), together with the incrementer proposed in \cite{gidney2015incr}, which runs in $\mathcal O(n)$. Since we have many dirty ancillae available in our recursive construction, the $n$-borrowed qubits incrementer in \cite{gidney2015incr} is sufficient: Using the ancilla-free adder by Takahashi \cite{takahashi2009quantum}, which requires no incoming carry, and its reverse to perform subtraction, one can perform the following sequence of operations to achieve an incrementer using $n$ borrowed ancilla qubits in an unknown initial state $\Ket g$:
\begin{align*}
	\Ket x\Ket g&\mapsto\Ket{x-g}\Ket{g}\\&\mapsto\Ket{x-g}\Ket{g'-1}
	\\&\mapsto\Ket{x-g-g'+1}\Ket{g'-1}\\&\mapsto\Ket{x+1}\Ket{g},
\end{align*}
where $g'$ denotes the two's-complement of $g$ and $g'-1=\overline g$, the bit-wise complement of $g$. A conditional incrementer can be constructed by either using two controlled adders as explained, or by applying an incrementer to a register containing both the target and control qubits of the conditional incrementer, where the control qubit is now the least significant bit \cite{gidney2015incr}. Then, the incrementer can be run on the larger register, followed by a final NOT gate acting on the control qubit (since it will always be toggled by the incrementer). In the latter version, one can either use one more dirty ancilla qubit for cases where $n\operatorname{mod}2=0$ or, alternatively, split the incrementer into two smaller ones as done in Ref.~\cite{gidney2015incr}. We will use the construction with an extra dirty qubit, since there are plenty of idle qubits available in Shor's algorithm.

\begin{figure}[t]
	\centering
	\includegraphics[width=.85\linewidth]{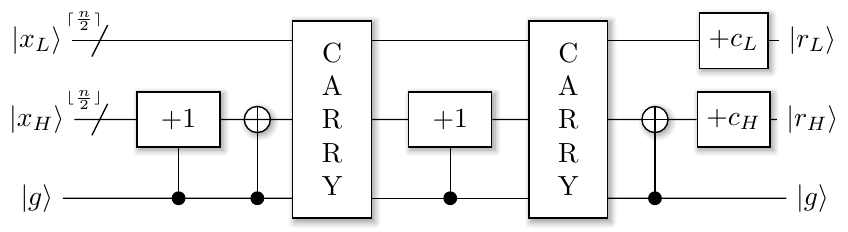}
	\bigskip
	\fcaption{The circuit of Fig. \ref{fig:addernlogn} for the case when the ancilla qubit is dirty (unknown initial state $\Ket g$, left unchanged by the computation).}
	\label{fig:addernlogndirty}
\end{figure}

In order to make the circuit depicted in Fig.~\ref{fig:addernlogn} work with a borrowed dirty qubit, the incrementer has to be run twice with a conditional inversion in between \cite{gidney2015incr}. The resulting circuit can be seen in Fig.~\ref{fig:addernlogndirty}. At the lowest recursion level, only 1-bit additions are performed, which can be implemented as a NOT gate on $x_i$ if $c_i=1$; all carries are accounted for earlier.

\subsection{Runtime analysis of the serial implementation}\label{sec:parallel}
In the serial version, we always reuse the one borrowed dirty ancilla qubit to hold the output of the CARRY gate, which is implemented as shown in Fig.~\ref{fig:adderex}. The CARRY gate has a Toffoli count of $T_{carry}(n)=4(n-2)+2$ (including the uncomputation of the ancilla qubits) and the controlled incrementer using $n$ borrowed dirty qubits features a Toffoli count of $T_{incr}(n)=2(2n-1)$ (2 additions). Both of these circuits have to be run twice on roughly $\frac n2$ qubits. Therefore, the first part of the recursion has a Toffoli count of $T_{rec}(n)=8n-8$. The recursion for the Toffoli count $T_{add}(n)$ of the entire addition circuit yields
\begin{align*}
	T_{add}(n) &= T_{add}\left(\left\lceil\frac n2\right\rceil\right) + T_{add}\left(\left\lfloor\frac n2\right\rfloor\right) + T_{rec}(n)\\
	&= 8n(\log_2n-2)+\mathcal O(1).
\end{align*}

For a controlled addition, only the two CNOT gates acting on the last bit in Fig.~\ref{fig:adderex} need to be turned into their controlled versions, which is another nice property of this construction.

\subsection{Parallel / Lower-depth version}
If the underlying hardware supports parallelization, one can compute the carries for the additions $+c_L$ and $+c_H$ in Fig.~\ref{fig:addernlogn} in parallel, at the cost of one extra qubit in state $\Ket0$ which will then hold the output of the CARRY computation of $+c_H$. Doing this recursively and noting that there must be at least two qubits of $x$ per CARRY gate, one sees that this circuit can be parallelized at a cost of $\frac n2$ ancilla qubits in state $\Ket 0$. Using the construction depicted in Fig.~\ref{fig:addernlogndirty} allows us to use $\frac n2$ borrowed dirty qubits instead. To see that this construction can be used in our implementation of Shor's algorithm, consider that during the modular multiplication 
\[
	\Ket x\Ket 0\mapsto\Ket x\Ket{(ax)\operatorname{mod}N}\;,
\]
we perform additions into the second register, conditioned on the output of the comparator in Fig.~\ref{fig:takahashi}. Therefore, $n$ qubits of the $x$-register are readily available to be used as borrowed dirty qubits, thus reducing the depth of our addition circuit to $\mathcal O(n)$.

Note that this is also possible if there is only one dirty ancilla available: Applying one round of the recursion in Fig.~\ref{fig:addernlogndirty} allows to run the low-depth version of $+c_L$ using $x_H$ as dirty qubits, before executing $+c_H$ using $r_L$ as dirty qubits.

\section{Modular multiplication}\label{sec:mult}
\begin{figure}[tb]
	\centering
	\includegraphics[width=.85\linewidth]{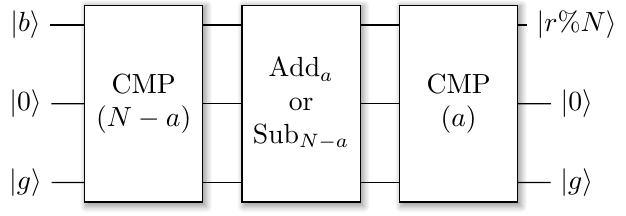}
	\bigskip
	\fcaption{Taken from \cite{takahashi2006quantum}: Construction of a modular adder $\Ket b\mapsto\Ket{r\operatorname{mod}N}$ with $r=a+b$, using a non-modular adder. The CMP gate compares the value in register $b$ to the classical value $N-a$, which we implement using our carry gate. The result indicates whether $b<N-a$, i.e., it indicates whether we must add $a$ or $a-N$. Finally, the indicator qubit is reset to $\Ket0$ using another comparison gate. In our implementation, the add/subtract operation uses between $1$ (serial) and $\frac n2$ (parallel) qubits of $g$.}
	\label{fig:takahashi}
\end{figure}

The modular multiplier can be constructed from a modular addition circuit using a repeated-addition-and-shift approach, as done in Refs.~\cite{beauregard2002circuit,takahashi2006quantum}:
\begin{align*}
(ax)\operatorname{mod}N &= (a(x_{n-1}2^{n-1}+\cdots+x_02^0))\operatorname{mod}N\\
	&= (((a2^{n-1})\operatorname{mod}N)x_{n-1}+\cdots+ ax_0)\;,
\end{align*}
where $x_{n-1},...,x_0$ is the binary expansion of $x$, and addition is carried out modulo $N$. Since $x_i\in\{0,1\}$, this can be viewed as modular additions of $(a2^i)\operatorname{mod}N$ conditioned on $x_i=1$. The transformation by Takahashi et al.~\cite{takahashi2006quantum} allows to construct an efficient modulo-$N$ addition circuit from a non-modular adder. For an illustration of the procedure see Fig.~\ref{fig:takahashi}, where the comparator can be implemented by applying our carry circuit on the inverted bits of $b$. Also, note that it is sufficient to turn the final CNOT gates (see Fig.~\ref{fig:adderex}) of the comparator in Fig.~\ref{fig:takahashi} into Toffoli gates in order to arrive at controlled modular addition, since the subsequent add/subtract operation is executed conditionally on the output of the comparator.

The repeated-addition-and-shift algorithm transforms the input registers \[
\Ket x\Ket 0\mapsto\Ket x\Ket{(a\cdot x)\operatorname{mod} N}\;.
\]
In Shor's algorithm, $2n$ such modular multiplications are required and in order to keep the total number of $2n+2$ qubits constant, the uncompute method from Ref.~\cite{beauregard2002circuit} can be used: After swapping the two $n$-qubit registers, one runs another modular multiplication circuit, but this time using subtraction instead of addition and with a new constant of multiplication, namely the inverse $a^{-1}$ of $a$ modulo $N$. This achieves the transformation
\begin{align*}
	\Ket x\Ket{(ax)\operatorname{mod}N}&\mapsto \Ket{(ax)\operatorname{mod}N}\Ket x\\
	&\mapsto\Ket{(ax)\operatorname{mod}N}\Ket{(x - a^{-1}ax)\operatorname{mod}N}
	\\&=\Ket{(ax)\operatorname{mod}N}\Ket0\;,
\end{align*}
as desired. In total, this procedure requires $2n+1$ qubits: $2n$ for the two registers and $1$ to achieve the modular addition depicted in Fig.~\ref{fig:takahashi}.

\section{Implementation and simulation results}\label{sec:exp}

In Shor's algorithm, a controlled modular multiplier is needed for the modular exponentiation which takes the form of a quantum phase estimation, since
\begin{align*}
	a^x\operatorname{mod}N &= a^{2^{n-1}x_{n-1}+2^{n-2}x_{n-2}\cdots+x_0}\operatorname{mod}N\\
	&=a^{2^{n-1}x_{n-1}}\cdot a^{2^{n-2}x_{n-2}}\cdots a^{x_0},
\end{align*}
where again $x_i\in\{0,1\}$ and multiplication is carried out modulo $N$. Thus, modular exponentiation can be achieved using modular multiplications by constants $\tilde a_i$ conditioned on $x_i=1$, where
\[
	\tilde a_i = a^{2^i}\operatorname{mod}N.
\] 
We do not have to condition our inner-most adders; we can get away with adding two controls to the comparator gates in Fig.~\ref{fig:takahashi}, which turns the CNOT gates acting on the last bit in Fig.~\ref{fig:adderex} into 3-qubit-controlled-NOT gates, which can be implemented using 4 Toffoli gates and one of the idle garbage qubits of $g$ \cite{barenco1995elementary}. Note that there are $n$ idle qubits available when performing the controlled addition/subtraction in Fig.~\ref{fig:takahashi} ($n-1$ qubits in $g$ plus the $x_i$ qubit the comparator was conditioned upon). The controlled addition/subtraction circuit can thus borrow $\frac n2$ dirty qubits from the $g$ register to achieve the parallelism mentioned in subsection~\ref{sec:parallel}, and the remaining $\frac n2$ dirty qubits can be used to decrease the depth of the incrementers in the recursive execution of the circuit in Fig.~\ref{fig:addernlogndirty}.

We implemented the controlled modular-multiplier performing the operation
\begin{align*}
	\Ket x\Ket 0&\mapsto\Ket x\Ket{(ax)\operatorname{mod}N}\\&\mapsto\Ket{(ax)\operatorname{mod}N}\Ket0
\end{align*}
in the LIQ$Ui\Ket{}$ quantum software architecture \cite{wecker2014liqui}. We extended LIQ$Ui\Ket{}$ by a reversible circuit simulator to enable large scale simulations of Toffoli based circuits. 

To test our circuit designs and gate estimates, we simulated our circuits on input sizes of up to $8,192$-bit numbers. The scaling results of the Toffoli count $T_{mult}(n)$ of our controlled modular-multiplier are as expected. 
Each of the two (controlled) multiplication circuits (namely compute/uncompute) use $n$ (doubly-controlled) modular additions. 
In addition to the comparators which require $\mathcal O(n)$ Toffoli gates, the modular addition circuit consists of two (controlled) addition circuits, which feature a Toffoli count of $T_{add}(n)=8n \log_2 n+\mathcal O(n)$.
Thus we have
\[
	T_{mult}(n) = 2n(2T_{add}(n) + \mathcal O(n)) = 32n^2 \log_2 n + \mathcal O(n^2)\;.
\]

\begin{figure}[t]
	\centering
	\resizebox{.7\linewidth}{!}{\input{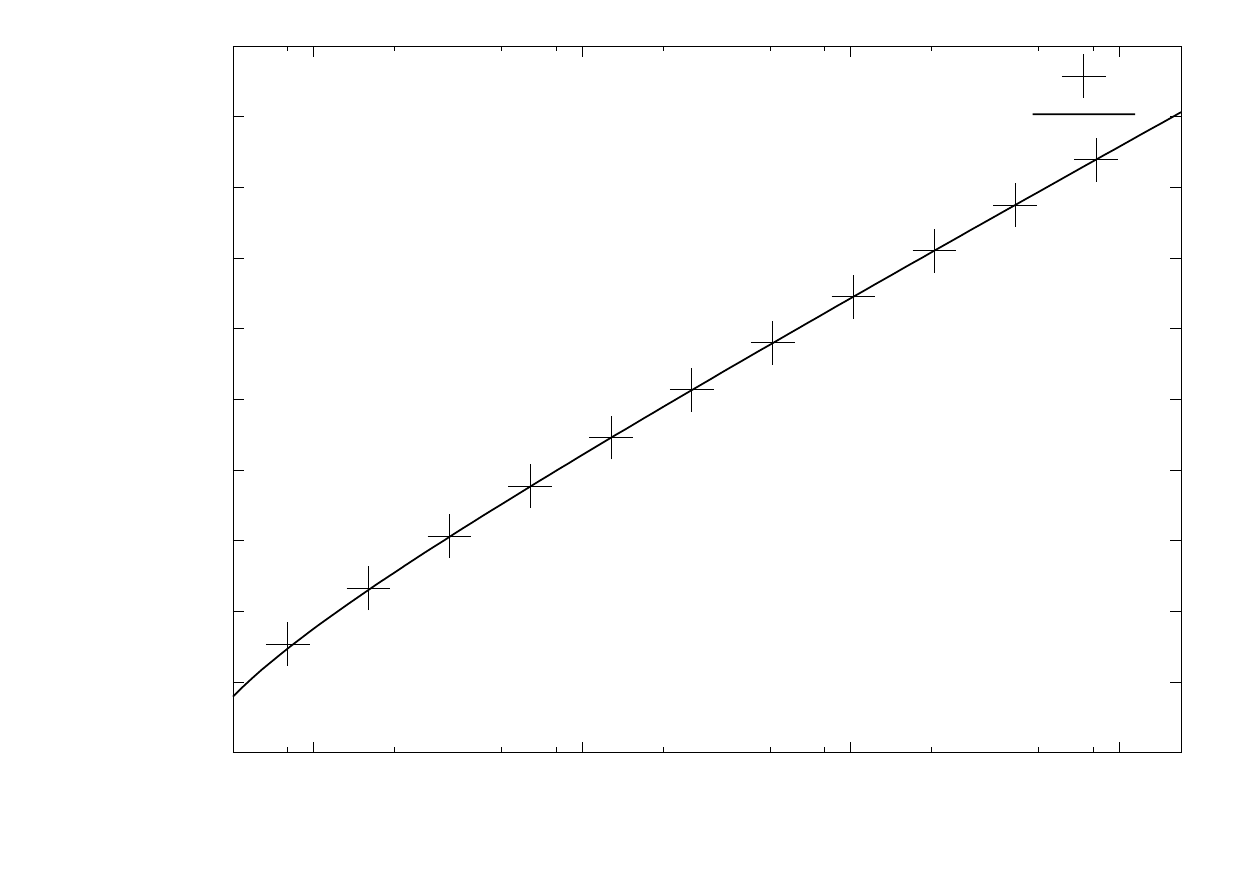}}
	\bigskip
	\fcaption{Scaling of the Toffoli count $T_M(n)$ with bit size $n$ for the controlled modular multiplier. Each data point represents a modular multiplication run (including uncompute of the $x$-register) using $n=2^m$ bits for each of the two registers, with $m\in\{3,...,13\}$.}
	\label{fig:scaling}
\end{figure}

The experimental data and the fit confirm this expected scaling, as shown in Fig.~\ref{fig:scaling}. 
Since $2n$ modular multiplications have to be carried out for an entire run of Shor's algorithm, the overall Toffoli count is 
\[
	T_{\text{Shor}}(n)=64n^3 \log_2 n + \mathcal O(n^3)\;.
\]

\section{Advantages of Toffoli circuits}\label{sec:advantages}
\subsection{Single-qubit rotation gate synthesis}
In order to apply a QEC scheme, arbitrary rotation gates have to be decomposed into sequences of gates from a (universal) discrete gate set -- a process called \textit{gate synthesis} -- for which an algorithm such as the ones in Refs.~~\cite{KMM:2016,Selinger:2015,RS:2016,BRS:2015,BRS:2015b} may be used. One example of a universal discrete gate set consists of the Pauli gates ($\sigma_x,\sigma_y,\sigma_z$), CNOT, Hadamard, and the T gate 
$\left(
\begin{smallmatrix}
1 & 0 \\ 0 & e^{i\pi/4}
\end{smallmatrix}\right)$. This synthesis implies a growth on the order of $\Theta(\log\frac 1\varepsilon)$ in the total number of gates, where $\varepsilon$ denotes the target precision of the synthesis.

In space-efficient implementations of Shor's algorithm by Beauregard~\cite{beauregard2002circuit} and Takahashi et al.~\cite{takahashi2006quantum}, the angles of the approximate QFT (AQFT) require synthesis. Since the approximation cutoff is introduced at $m\in\Theta(\log n)$, the smallest Fourier angles are $\theta_m\in\Theta(\frac 1{2^m})$ \cite{barenco1996approximate}. Hence, the target precision of the synthesis can be estimated to be in $\omega(\frac 1n)$ and, thus, the overall gate count and depth of the previous circuits by Takahashi et al.~and Beauregard are in $\Theta(n^3\log^2n)$ and $\Theta(n^3\log n)$, respectively.

Toffoli based networks, on the other hand, do not suffer from synthesis overhead. A Toffoli gate can be decomposed exactly into Clifford and T gates using 7 T-gates, or less if global phases can be ignored~\cite{barenco1995elementary,jones2013toffoli}. While the rotation gates from the semi-classical inverse QFT (see Fig.~\ref{fig:shor}) require synthesis, this does not affect the asymptotic scaling. Therefore, the overall gate count and depth of our circuit remain in $\Theta(n^3\log n)$ and $\Theta(n^3)$, respectively.

\subsection{Design for testability}

In classical computing, thoroughly tested hardware and software components are preferred over the ones that are not, especially for applications where system-failure could have catastrophic effects. The same may be true for quantum computing: Both software and hardware will need to be tested in order to guarantee the correctness of each and every component involved in a computation for building large circuits such as the ones used for factoring using Shor's algorithm. While a full functional simulation may be possible for arbitrary circuits up to almost 50 qubits with high-performance simulators run on supercomputers~\cite{haner2016high,smelyanskiy2016qhipster}, simulating a moderately-sized future quantum computer with just 100 qubits is not feasible on a (future) classical computer due to the exponential scaling of the required resources. For Toffoli networks, on the other hand, classical reversible simulators can be used, which run the circuit on a computational basis state and only update one single state for each gate. This enables thorough testing of logical level circuits such as the modular multiplication circuit presented in this paper. 

Furthermore, when Toffoli networks are run on actual quantum hardware, the circuits can be debugged efficiently and faults can be localized using binary search. Faults here include missing-gate faults (which is similar to a classical ``stuck-at fault'' \cite{EDA}) and errors in the computational basis such as, e.g., bitflips. The idea is to run the actual physical implementation of the network (followed by final measurement in the computational basis) on a sample set of basis states which serve as test vectors to trigger the faults. As the distribution under correct operation is always close to a delta function in total variation distance, we have an efficient test if an error occurred in the entire network. If so, one can subdivide the Toffoli networks in two parts and recursively apply the procedure to both parts, using as input the ideal vectors obtained from the logical level simulation. Eventually this will lead to localization of all faults that might be present in the implementation, provided that the choice of the initial sample set triggers all faults that might be present in the network. Note that this kind of debugging would not be possible for, e.g., QFT-based addition circuits, as intermediate states might be in superposition of exponentially many basis states.

\section{Summary and Outlook}

We present a Toffoli based in-place addition circuit which can be used to implement Shor's algorithm using $2n+2$ qubits. Our implementation features a size in $\mathcal O(n^3\log n)$, and a depth in $\mathcal O(n^3)$. In contrast to previous space-efficient implementations \cite{beauregard2002circuit,takahashi2006quantum}, our modular multiplication circuit only consists of Toffoli and Clifford gates. In addition to facilitating the process of debugging future implementations, having a Toffoli based circuit also eliminates the need for single-qubit-rotation synthesis when employing quantum error-correction. This results in a better scaling of both size and depth by a factor in $\Theta(\log n)$.

Our main technical innovation is the implementation of an addition by a constant that can be performed in $\mathcal O(n\log n)$ operations and that uses between $1$ and $n$ ancillas, all of which can be dirty, i.e., can be taken from other parts of the computation that are currently idle.

As mentioned in \cite{takahashi2006quantum}, it would be interesting to see whether one can find a linear-time constant-adder which does not require $\Theta(n)$ clean ancilla qubits. This would allow to decrease the size of our circuit to its current depth of $\mathcal O(n^3)$ without having to increase the total number of qubits to $3n+2$. Also, similar to \cite{vanMeter2005modular,kutin2006shor}, it would be interesting to find implementations of Shor's algorithm that are geometrically constrained but yet make use of dirty ancillas to reduce the overall number of qubits required. 

\section*{Acknowledgments}
We thank Alan Geller, Craig Gidney, Dave Wecker, and Nathan Wiebe for helpful discussions.


\bibliographystyle{unsrt}
\bibliography{references}
\end{document}